\documentclass[aps,prc,twocolumn,superscriptaddress,showpacs,floatfix]{revtex4}
\usepackage{epsfig}
\usepackage{dcolumn}
\begin{document}

\def\GSI{Gesellschaft f\"ur Schwerionenforschung mbH, D-64291 Darmstadt,
Germany}
\def\GANIL{GANIL, CEA et IN2P3-CNRS, F-14076 Caen, France}
\def\IPNO{Institut de Physique Nucl\'eaire, IN2P3-CNRS et Universit\'e, F-91406
Orsay, France}
\def\LPC{LPC, IN2P3-CNRS, ISMRA et Universit\'e, F-14050 Caen, France}
\def\SACLAY{DAPNIA/SPhN, CEA/Saclay, F-91191 Gif sur Yvette, France}
\def\LYON{Institut de Physique Nucl\'eaire, IN2P3-CNRS et Universit\'e, F-69622
Villeurbanne, France}
\def\NAPOLI{Dipartimento di Scienze Fisiche e Sezione INFN, Univ. Federico II,
I-80126 Napoli, Italy}
\def\CATANIA{Dipartimento di Fisica dell' Universit\`a and INFN, I-95129
Catania, Italy}
\def\WARSAW{A.~So\l{}tan Institute for Nuclear Studies, Pl-00681 Warsaw, Poland}
\def\MOSCOW{Institute for Nuclear Research, 117312 Moscow, Russia}
\def\IFJ{H. Niewodnicza\'nski Institute of Nuclear Physics, Pl-31342 Krak\'ow,
Poland}
\def\CNAM{Conservatoire National des Arts et M\'etiers, F75141 Paris cedex 03, 
France}

\title{Fragmentation in Peripheral Heavy-Ion Collisions:
from Neck Emission to\\ Spectator Decays}

\affiliation{\GSI}
\affiliation{\GANIL}
\affiliation{\IPNO}
\affiliation{\LPC}
\affiliation{\SACLAY}
\affiliation{\LYON}
\affiliation{\NAPOLI}
\affiliation{\CATANIA}
\affiliation{\WARSAW}
\affiliation{\IFJ}
\affiliation{\CNAM}
\affiliation{\MOSCOW}

\author{J.~{\L}ukasik}		\affiliation{\GSI}\affiliation{\IFJ}
\author{G.~Auger}		\affiliation{\GANIL}
\author{M.L.~Begemann-Blaich}	\affiliation{\GSI}
\author{N.~Bellaize}		\affiliation{\LPC}
\author{R.~Bittiger}		\affiliation{\GSI}
\author{F.~Bocage}		\affiliation{\LPC}
\author{B.~Borderie}		\affiliation{\IPNO}
\author{R.~Bougault}		\affiliation{\LPC}
\author{B.~Bouriquet}		\affiliation{\GANIL}
\author{J.L.~Charvet}		\affiliation{\SACLAY}
\author{A.~Chbihi}		\affiliation{\GANIL}
\author{R.~Dayras}		\affiliation{\SACLAY}
\author{D.~Durand}		\affiliation{\LPC}
\author{J.D.~Frankland}		\affiliation{\GANIL}
\author{E.~Galichet}		\affiliation{\IPNO}\affiliation{\CNAM}
\author{D.~Gourio}		\affiliation{\GSI}
\author{D.~Guinet}		\affiliation{\LYON}
\author{S.~Hudan}		\affiliation{\GANIL}
\author{B.~Hurst}		\affiliation{\LPC}
\author{P.~Lautesse}		\affiliation{\LYON}
\author{F.~Lavaud}		\affiliation{\IPNO}
\author{A.~Le~F\`evre}		\affiliation{\GSI}
\author{R.~Legrain}		\altaffiliation[]{deceased}\affiliation{\SACLAY}
\author{O.~Lopez}		\affiliation{\LPC}
\author{U.~Lynen}		\affiliation{\GSI}
\author{W.F.J.~M\"uller}	\affiliation{\GSI}
\author{L.~Nalpas}		\affiliation{\SACLAY}
\author{H.~Orth}		\affiliation{\GSI}
\author{E.~Plagnol}		\affiliation{\IPNO}
\author{E.~Rosato}		\affiliation{\NAPOLI}
\author{A.~Saija}		\affiliation{\CATANIA}
\author{C.~Schwarz}		\affiliation{\GSI}
\author{C.~Sfienti}		\affiliation{\GSI}
\author{J.C.~Steckmeyer}	\affiliation{\LPC}
\author{B.~Tamain}		\affiliation{\LPC}
\author{W.~Trautmann}		\affiliation{\GSI}
\author{A.~Trzci\'{n}ski}	\affiliation{\WARSAW}
\author{K.~Turz\'o}		\affiliation{\GSI}
\author{E.~Vient}		\affiliation{\LPC}
\author{M.~Vigilante}		\affiliation{\NAPOLI}
\author{C.~Volant}		\affiliation{\SACLAY}
\author{B.~Zwiegli\'{n}ski}	\affiliation{\WARSAW}
\author{A.S.~Botvina}		\affiliation{\GSI}\affiliation{\MOSCOW}
\collaboration{The INDRA and ALADIN Collaborations}
\noaffiliation

\date{\today}

\begin{abstract}

Invariant cross sections of intermediate mass fragments in peripheral
collisions of $^{197}$Au on $^{197}$Au  at incident energies between 40 and 150
MeV per nucleon  have been measured with the 4$\pi$ multi-detector INDRA.  The
maximum of the fragment production is located near mid-rapidity at the lower
energies and moves gradually  towards the projectile and target rapidities as
the energy is increased. Schematic calculations within an extended Goldhaber
model suggest that the observed cross-section distributions and their evolution
with energy are predominantly the result of the clustering requirement for 
the emerging
fragments and of their Coulomb repulsion  from the projectile and target
residues. The quantitative comparison with transverse energy spectra and
fragment  charge distributions emphasizes the role of hard scattered nucleons
in the fragmentation process.

\end{abstract}


\maketitle

The production of nuclear fragments is a decay mode that persists up to the 
highest bombarding energies in heavy-ion collisions. Its characteristics,
however, are considerably different in the Fermi energy  regime and in the
relativistic regime. At bombarding energies near the  Fermi energy, the
fragment distributions are concentrated at mid-rapidity, neck breaking 
mechanisms
have been suggested by various transport simulations, and strong Coulomb
effects are generated by  the presence of slowly moving heavy residues
\cite{mont94, toke95, colo95, demp96, luka97, nebau99, plagnol00, dore01, 
poggi01, smm2, piant02, milazzo02, baran02, davin02}.

At relativistic energies, the fragment production in peripheral collisions is
concentrated at  projectile and target rapidities 
\cite{hubele92, hauger96, xi97} 
with invariance properties that suggest a high degree  of equilibration for
the decaying spectator systems \cite{schuett96, gait00, insolia00}. Because of
their prevailing statistical character these fragmentations are often seen
distinguished from the dynamical processes at intermediate energies.

The present study is directed to the transition region between these  seemingly
so different regimes with the aim to identify the basic  ingredients that
govern the evolution of the fragment production  mechanisms. As the main
result, we find that the same criteria are at work throughout the covered
energy range and that, of these, the clustering (coalescence) criterion is
essential for producing the observed features.

The experiments were performed at the GSI Darmstadt with beams of $^{197}$Au,
delivered by the heavy-ion synchrotron SIS and directed onto $^{197}$Au
targets of 2-mg/cm$^2$ areal thickness placed inside the INDRA multidetector
\cite{pouthas}. The incident energies varied between 40 to 150 MeV per
nucleon, a range of energies that corresponds to relative velocities between
once and twice the Fermi value. Further details of the experimental and
calibration procedures  may be found in Refs. \cite{luka02, trzc02}.

As described previously, the total  transverse energy $E^{12}_{\perp}$ of light
charged particles ($Z \le$ 2) has been used as an impact parameter selector
\cite{luka02}. The measured $E^{12}_{\perp}$ spectra scale linearly with the 
incident energy, and the scaled spectra coincide. The relation  between
$E^{12}_{\perp}$ and the reduced impact parameter $b/b_{\rm max}$ has been
obtained by applying the geometrical prescription of Ref.~\cite{cavata90}. It
is linear in very good approximation, where $b/b_{\rm max}$ decreases with
increasing $E^{12}_{\perp}$. Eight finite impact parameter bins were generated,
with the most central bin 8 covering impact parameters up to 5\% of $b_{\rm
max}$. The remaining part of the $E^{12}_{\perp}$ spectrum has been divided
into 7 bins of equal width, corresponding to 7 bins of  approximately equal
width in $b$, of which bin 1 contains the most peripheral events. An event was
registered if at least three detectors had recorded a hit.

\begin{figure}[!htb]
     \epsfxsize=\columnwidth
     \centerline{\epsffile{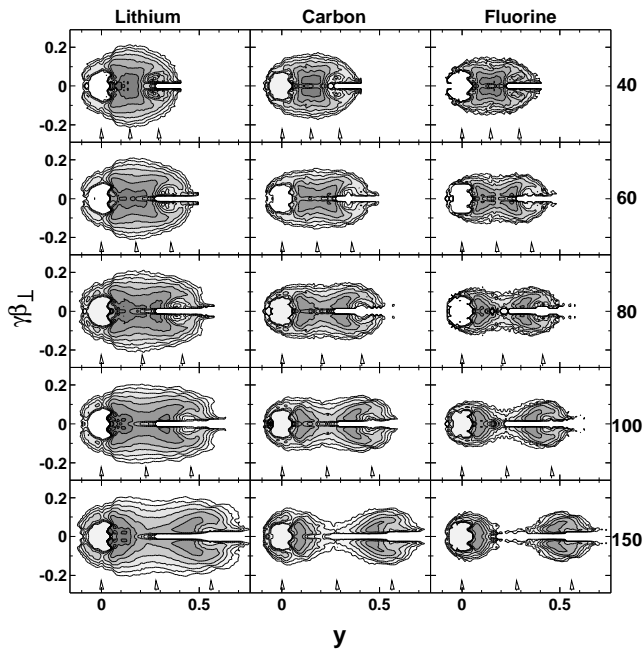}}

     \caption{Invariant cross section distributions as a function of
     transverse velocity  $\gamma\beta_{\perp}$ and rapidity $y$ for 
     fragments with $Z$ = 3, 6, 9 from the  most peripheral collisions 
     (bin 1) of $^{197}$Au + $^{197}$Au at $E/A$ =  40, 60, 80, 100 and 
     150 MeV, as indicated. The cross sections are normalized relative to
     each other, separately for each fragment species. Near the target 
     rapidity they are affected by the thresholds for fragment
     identification. The arrows indicate the target, center-of-mass and 
     projectile rapidities (from left to right).}
\label{fig:au15}
\end{figure}
 
Invariant cross section distributions for lithium, carbon, and fluorine
fragments emitted in peripheral $^{197}$Au + $^{197}$Au collisions are shown
in Fig.~\ref{fig:au15}.
They illustrate the characteristic evolution of the emission pattern
as the incident energy is raised from 40 up to 150 MeV per nucleon.

The Coulomb repulsion from the projectile and target residues that survive the
collision at peripheral impact parameters is evident in all
distributions. The distribution of intermediate-mass fragments, however, is not
isotropic with respect to these residues. They are emitted  with a preference
toward mid-rapidity which produces a pronounced forward-backward asymmetry 
relative to the rapidities of the projectile and target residues. The resulting
cross-section maximum at mid-rapidity at the lower energies gives way to a
saddle-type structure at mid-rapidity at the higher energies. 
Here the fragment production is
increasingly concentrated at the projectile and  target rapidities, a trend
that is more pronounced at even higher incident 
energies and generally characteristic for the
relativistic regime \cite{schuett96}. At a given energy, the probability for
forming fragments at mid-rapidity decreases rapidly with increasing $Z$.

The kinetic properties of the mid-rapidity fragments in peripheral collisions,
in spite of this apparent evolution, have been found rather constant throughout
the covered range of projectile energies \cite{luka02}. The transverse
velocities are invariant with respect to both, the incident energy and the 
fragment $Z$. This property and the large value 
$\langle E_{\rm t} \rangle \approx$
30 MeV of the  corresponding mean transverse energies were suggested to be the
result of  the combined action of the intrinsic Fermi motion, the acquired
transverse  energies of scattered nucleons and of the Coulomb repulsion from
the heavy  residues.

To obtain more quantitative predictions, a Monte-Carlo procedure based on these
considerations has been developed. The original idea of Goldhaber
\cite{gold74} of taking the fragment momentum as the sum of the momenta of the 
constituent nucleons picked randomly from a single Fermi sphere has been 
extended by including also the Pauli-allowed distribution of hard scattered
nucleons. The corresponding calculations proceed in three main steps:

(i) The momenta of the $A$ nucleons of a fragment with mass number $A$  are
randomly picked from 2 Fermi spheres ($p_{\rm F}$ = 265 MeV/c), 
separated by a relative momentum per
nucleon calculated at the distance of closest approach of the two nuclei
colliding at a given impact parameter, and also from the momentum distribution
generated in hard two-body collisions of nucleons from the projectile and
target. Isotropic scattering has been assumed for these  
collisions, and they were requested to obey the Pauli principle. 
The Pauli-allowed  momentum space changes with increasing incident
energy and widens both toward  lower and higher transverse momenta. In the
calculations, a fixed impact parameter of 10 fm was used, and the number of
hard scattered nucleons to be incorporated in a fragment was taken as a
parameter. The results presented here were obtained assuming a binomial
distribution of the number of hard scattered nucleons in a fragment with
mean values that were found to be rather small (see below).

\begin{figure}[!htb]
     \epsfxsize=0.9\columnwidth
     \centerline{\epsffile{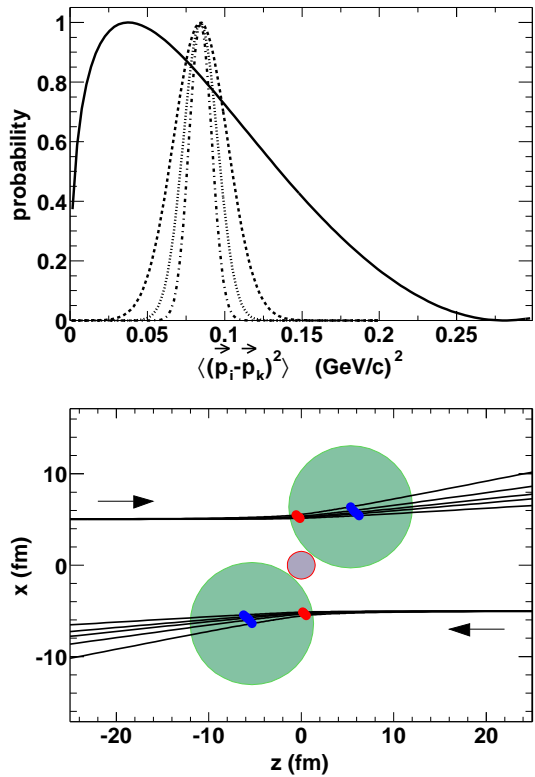}}

     \caption{Top panel: Probability distributions of the mean squared relative
     momentum of nucleon pairs picked randomly from a single Fermi sphere for
     fragments with $A$ = 2 (solid line) and Gaussian approximations of the
     distributions for $A$ = 7, 14 and 28 (dashed, dotted and dash-dotted
     lines, respectively). Bottom panel: Initial spatial conditions for Coulomb
     trajectory  calculations in the reaction plane, $x$ vs $z$. The lines
     represent the trajectories of the projectile and the target at incident
     energies $E/A$ = 40 to 150 MeV and for a fixed impact parameter of 10 fm.
     The dots placed on the trajectories mark the positions of the centers of
     the nuclei at the distance of closest approach (gray) and in the 
     exit-channel sticking configuration (black). The arrows indicate the
     directions of the ion motion.}

\label{fig:model}
\end{figure}

(ii) A clustering criterion has been implemented by randomly accepting or
rejecting the trial clusters according to the probability distribution of the
mean squared relative momentum, $\sum_{i\neq
j}(\vec{p}_i-\vec{p}_j)^2/(A(A-1))$, with that of $A$ nucleons drawn  randomly
from a single Fermi sphere of the radius of $p_{\rm F}$ = 265 MeV/c.  Examples
of these reference distributions are shown in the upper part of 
Fig.~\ref{fig:model} 
for fragments containing 2, 7, 14, and 28 nucleons. They are
characterized by a mean value $\langle p_{\rm rel}^2 \rangle$ and a  variance
$\sigma^2 (p_{\rm rel}^2)$ given by:
\[
\langle p_{\rm rel}^2 \rangle = \frac{6}{5} p_{\rm F}^{2}; \;\;\;
\sigma^2 (p_{\rm rel}^2) = \frac{4}{25} p_{\rm F}^{4} \frac{12 A+30}{7A(A-1)}.
\]
The mean value of the distributions is constant while the variances decrease
with increasing fragment mass $A$. The asymmetry apparent for lighter fragments
disappears for heavier ones and, for $A \ge$ 7, the distributions were 
approximated by Gaussians. The clustering criterion strongly suppresses the
formation of fragments with large relative momenta of constituent nucleons.

The fragments selected by random picking of the momenta from a single Fermi
sphere are not entirely cold (see e.g. \cite{murphy}), and neither 
are the fragments defined with the above criterion. 
Temperatures deduced by fitting a Fermi
distribution to the distribution of nucleon momenta in the center of mass of a
fragment drops, approximately like $1/\sqrt A$, from about 5 MeV for $A$ = 10 
to near 0 for very large clusters. 
The sequential decay of excited clusters is expected to modify
the fragment yields and kinetic energies, but not to
an extent that would substantially modify the present conclusions.

\begin{figure}[!htb]
     \epsfxsize=0.7\columnwidth
     \centerline{\epsffile{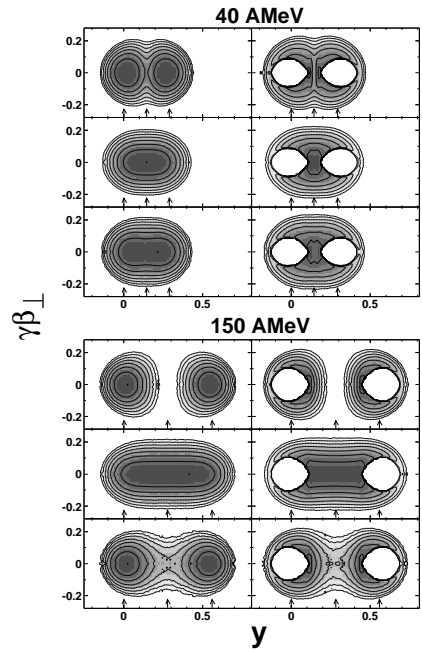}}

     \caption{Invariant cross sections of $^{7}$Li fragments as a function of
     transverse velocity $\gamma\beta_{\perp}$ and rapidity $y$,
     as obtained from
     the  model calculations for peripheral $^{197}$Au + $^{197}$Au collisions
     at $E/A$ = 40 MeV (top part)  and 150 MeV (bottom). The left and
     the right columns show the initial and asymptotic distributions, i.e.
     before and after the Coulomb trajectory calculations, respectively. The
     results without nucleon mixing, without the clustering criterion, and of
     the full calculation are shown in the top, middle, and bottom rows for
     both energies, respectively. The arrows have the same meaning as in 
     Fig.~\protect\ref{fig:au15}.}

\label{fig:k34}
\end{figure}

The specific clustering criterion is adapted from the original
Goldhaber formulation of the fragmentation problem by generalizing it
from a single
Fermi sphere to the three distributions considered here. 
It differs from standard coalescence criteria based on a fixed momentum
radius and reduces
the number of free parameters of the model as it depends only on $p_{\rm F}$.

(iii) Accepted fragments have been placed, in coordinate space, in between the
two  residues in an aligned sticking configuration which served as the 
starting point for 3-body Coulomb trajectory calculations  
(Fig.~\ref{fig:model}, bottom panel). The initial momentum of the produced
fragment is determined by steps (i) and (ii), while the starting momenta
of the heavy fragments were generated with two-body trajectory calculations 
for the entrance channel, using Coulomb and proximity forces \cite{prox}. The
residue trajectories remain virtually undisturbed in the exit channel.
This simplified reaction picture assumes that the fragment is formed in the
contact zone of the colliding nuclei and released as soon as the
remaining residues separate by more than the diameter of the fragment which is
assumed to be spherical. The effect of additional charged particles is 
ignored since their multiplicity in peripheral collisions is small.

The results obtained with this procedure for 40 and 150 MeV per nucleon, the
lowest and  highest bombarding energy included in this study, are shown in 
Fig.~\ref{fig:k34} for the case of fragments with mass number $A$ = 7. Three
different conditions were chosen for the calculations which all have 
the assumption in common that the number of hard scattered nucleons
in a fragment is given by a binomial distribution with the mean value of 1.
The resulting cross section distributions are rather similar at the lowest 
energy but become distinctively different as the energy is raised.

\begin{figure}[!htb]
     \epsfxsize=\columnwidth
     \centerline{\epsffile{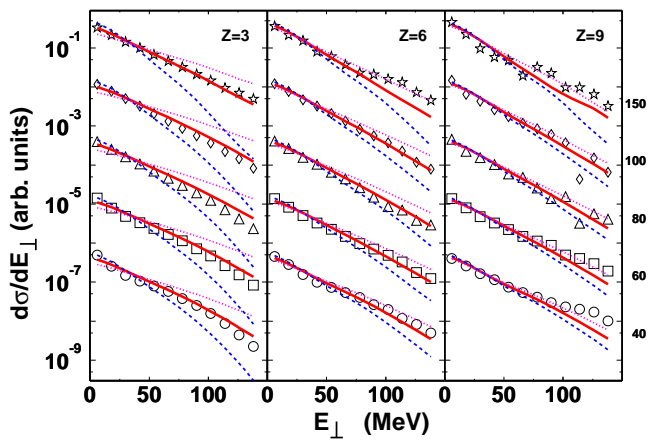}}

     \caption{Experimental (symbols) and calculated (lines) transverse-energy
     spectra of $^{7}$Li (left panel), $^{13}$C (middle), and 
     $^{19}$F fragments 
     (right) emitted at mid-rapidity in peripheral $^{197}$Au + $^{197}$Au
     collisions at five incident energies, as indicated. Solid, dashed and
     dotted lines correspond to the calculations with on average one, zero or
     two scattered nucleons in a fragment, respectively. 
     In each panel, the experimental spectra are displaced vertically 
     by consecutive factors of 30, and the calculated spectra are individually
     normalized relative to the corresponding measurement.}
\label{fig:ene}      
\end{figure}

If all the non-scattered nucleons forming a fragment are exclusively taken
from only one of  the collision partners, the mid-rapidity region cannot be
populated at the higher bombarding energies (top row of Fig.~\ref{fig:k34} for
150 AMeV). At the lower energies (top row for 40 AMeV) these  evaporation-like
events are copiously found at mid-rapidity because the Coulomb maxima, forming
the characteristic rings in the two-dimensional representation, intersect. 
If the clustering criterion is dropped,
(middle rows for both energies), the fragment distributions at high incident
energies populate nearly homogeneously the full range  from projectile to
target rapidities, again in contrast to the observation (cf. 
Fig.~\ref{fig:au15}).
Both ingredients, the mixing of target and projectile nucleons 
(see also \cite{nebau99}) and the clustering criterion, are indispensable 
if more realistic distributions are to be generated that resemble those 
observed (bottom rows for both energies). 
The Coulomb field and its interplay with the
initial momenta has a strong effect of rearranging and focusing the asymptotic
fragments (cf. Ref.~\cite{piant02}). Their distribution extends to the
projectile and target rapidities even though all fragments are initially placed
in the aligned sticking configuration in between the two residues (see (iii)
above). These specific initial conditions naturally produce the
forward-backward asymmetry with respect to the heavy-residue rapidities.

For a more quantitative comparison, measured and calculated transverse-energy 
spectra for mid-rapidity 
fragments with $Z$ = 3, 6, 9 are shown in Fig.~\ref{fig:ene}. 
Bins centered at mid-rapidity with widths of
25\% of the projectile rapidity $y_{\rm p}$ were chosen for the selection.
The exponential spectra shapes correspond
to normal statistical emissions.
Only the high value of the mean transverse energy ($\approx$ 30 MeV) has 
been found difficult to be explained within thermal models, 
as discussed in Ref. \cite{luka02}. 
The shapes of these spectra,
including their invariance with bombarding energy and $Z$,
are rather well reproduced 
if on average one scattered nucleon from the Pauli-allowed
distribution of hard scattered nucleons is included in the fragment 
(solid lines). 
If scattered nucleons are excluded (dashed lines) or if their mean number 
is increased to 2 (dotted lines), the resulting spectra deviate significantly
from the experiment, in particular for the lighter fragments.


A small number of scattered nucleons in a fragment is
consistent with the small number of primary nucleon-nucleon collisions 
expected for the very peripheral collisions selected with bin 1. 
The increasingly harder transverse-energy spectra in more central collisions
\cite{luka02} require more scattered nucleons to be included, in particular 
at the higher incident energies. For 100 and 150 MeV per nucleon, the lithium
spectra of bins 2 and 3, e.g., are well described if their mean number 
in a fragment is increased to 1.5 and 2, respectively. 

\begin{figure}[!htb]
     \epsfxsize=\columnwidth
     \centerline{\epsffile{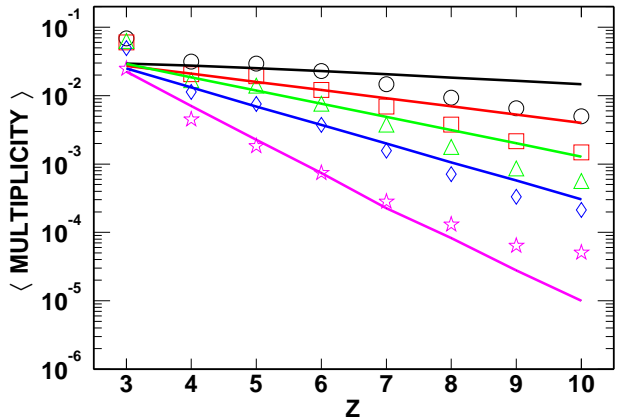}}

     \caption{Experimental (symbols) and calculated (lines) mean multiplicity
     of mid-rapidity fragments in peripheral $^{197}$Au + $^{197}$Au collisions
     (bin 1) as a function of Z. 
     The calculated spectra are normalized at $Z=6$,
     separately for each of the five bombarding energies $E/A$ = 
     40, 60, 80, 100, 150 MeV (from top to bottom).}

\label{fig:z}     
\end{figure} 

The general trends of the experimental cross section distributions as a 
function of the fragment $Z$ are also rather well reproduced. 
This is demonstrated in Fig.~\ref{fig:z}
for the range of mid-rapidity fragments with 3$\leq$Z$\leq$10. 
The $Z$ distributions are approximately exponential 
functions with slopes that increase monotonically with 
the incident energy. The calculations have been performed
for masses $A=2Z+1$ and for, on average, one scattered nucleon 
in a fragment. 
The $Z$ dependence of the differential multiplicities is nearly perfectly
reproduced for 100 MeV per nucleon
but the variation of the slopes with incident energy 
is slightly overpredicted. With the chosen normalization at $Z=6$, 
this has the 
effect that the yields of the heavier fragments are underestimated at the 
highest and overestimated at the lower bombarding energies. The yields of
lithium fragments are also too low by about a factor of two at all 
energies except 150 MeV per nucleon, where the measured yields are lower. 
Overall, however, the agreement is very satisfactory.

From the good qualitative (Figs. \ref{fig:au15} and \ref{fig:k34})
and quantitative agreement (Figs. \ref{fig:ene} and \ref{fig:z}) 
we may thus conclude  that the
presented extended Goldhaber scenario contains the essential ingredients of the
fragmentation mechanism in peripheral collisions. Of these, the clustering
criterion is responsible for the evolution of the cross section distributions
from mid-rapidity to the spectator rapidities with increasing projectile
energy. It is furthermore important to permit  the mixing of projectile and
target nucleons and to  include hard-scattered nucleons from Pauli-allowed
collisions into the  produced fragments. This latter observation even suggests
the hypothesis that the hard-scattered nucleons may 
be required for initiating 
the fragment production at mid-rapidity in peripheral collisions. 
The correct prediction of the role of scattered nucleons may therefore 
be considered as an important test for any dynamical model to describe the
mid-rapidity emission. 
It is, finally, remarkable how far the fragment
distributions extend into the projectile and target rapidity regions, as a
consequence of the combined action of the initial momenta and of the Coulomb
forces giving rise to the characteristic ring structures. Apparently, 
fragments observed there do not necessarily have to be considered as 
emitted from the excited residues but may to a large part originate 
from the contact zone, or neck region, formed during the reaction.

The authors would like to thank the staff of the GSI for providing 
high-quality $^{197}$Au beams and for technical support. M.B. and C.Sc.
acknowledge the financial support of the Deutsche Forschungsgemeinschaft under
the Contract Nos. Be1634/1 and Schw510/2-1, respectively; D.Go. and C.Sf.
acknowledge the receipt of  Alexander-von-Humboldt fellowships. This work was
supported by the European Community under Contract No. ERBFMG\-ECT\-950083.


\begin{thebibliography}{99}

\bibitem{mont94}
C.P.~Montoya {\it et al.},
Phys. Rev. Lett. {\bf 73}, 3070 (1994).

\bibitem{toke95}
J.~T\~oke {\it et al.},
Phys. Rev. Lett. {\bf 75}, 2920 (1995).

\bibitem{colo95}
M.~Colonna {\it et al.},
Nucl.~Phys. {\bf A589}, 160 (1995).

\bibitem{demp96} 
J.F.~Dempsey {\it et al.},
Phys. Rev. C {\bf 54}, 1710 (1996).

\bibitem{luka97}
J.~\L{}ukasik {\it et al.},
Phys. Rev. C {\bf 55}, 1906 (1997).

\bibitem{nebau99}
R.~Nebauer and J.~Aichelin,
Nucl. Phys. {\bf A650}, 65 (1999).

\bibitem{plagnol00} 
E.~Plagnol {\it et al.},
Phys. Rev. C {\bf 61}, 014606 (2000).

\bibitem{dore01}
D.~Dor\'e {\it et al.},
Phys. Rev. C {\bf 63}, 034612 (2001).

\bibitem{poggi01}
G.~Poggi,
Nucl.~Phys. {\bf A685}, 296c (2001).

\bibitem{smm2} 
A.S.~Botvina and I.N.~Mishustin, Phys. Rev. C {\bf63}, 061601(R) (2001).

\bibitem{piant02}
S. Piantelli {\it et al.},
Phys. Rev. Lett. {\bf 88}, 052701 (2002).

\bibitem{milazzo02}
P. M. Milazzo {\it et al.}, Nucl.~Phys. {\bf A703}, 466 (2002).
       
\bibitem{baran02}
V.~Baran {\it et al.}, Nucl.~Phys. {\bf A703}, 603 (2002).

\bibitem{davin02}
B.~Davin {\it et al.}, Phys. Rev. C {\bf 65}, 064614 (2002).

\bibitem{hubele92}
J. Hubele {\it et al.},
Phys. Rev. C {\bf 46}, R1577 (1992).

\bibitem{hauger96}
J.A.~Hauger {\it et al.},
Phys. Rev. Lett. {\bf 77}, 235 (1996).

\bibitem{xi97}
Hongfei~Xi {\it et al.},
Z.~Phys.~A {\bf 359}, 397 (1997);
Eur. Phys. J. A {\bf 1}, 235 (1998)
\
\bibitem{schuett96}
A. Sch\"uttauf {\it et al.},
Nucl. Phys. {\bf A607}, 457 (1996).

\bibitem{gait00}
T. Gaitanos {\it et al.},
Phys. Lett. B {\bf 478}, 79 (2000).

\bibitem{insolia00}
A.~Insolia {\it et al.},
Phys. Rev. C {\bf 61}, 044902 (2000).

\bibitem{pouthas} 
J. Pouthas {\it et al.}, 
Nucl. Instr. Meth. in Phys. Res. {\bf A357}, 418 (1995).

\bibitem{luka02}
J.~\L{}ukasik {\it et al.},
preprint GSI 2002-22, nucl-ex/0207015 (2002), 
Phys. Rev. C {\bf 66}, 064606 (2002).

\bibitem{trzc02}
A.~Trzci\'{n}ski {\it et al.},
GSI preprint (2002).	

\bibitem{cavata90}
C. Cavata {\it et al.}, 
Phys. Rev. C {\bf 42}, 1760 (1990).

\bibitem{gold74}
A.S.~Goldhaber,
Phys.~Lett.~{\bf 53B}, 306 (1974).

\bibitem{murphy}
M.J.~Murphy,
Phys.~Lett.~{\bf 135B}, 25 (1984).

\bibitem{prox}
C.~Ng\^{o} {\it et al.},
Nucl. Phys. {\bf A252}, 237 (1975).


\end{thebibliography}
\end{document}